\documentclass[preprint,prd,aps,amssymb]{revtex4}

\usepackage{graphicx}
\usepackage{axodraw}

\newcommand{\be}{\begin{equation}}
\newcommand{\ene}{\end{equation}}
\newcommand{\bea}{\begin{eqnarray}}
\newcommand{\enea}{\end{eqnarray}}

\begin{document}

\title {\large Lepton Flavor Violation in the SUSY-GUT 
Models \\ with Lopsided Mass Matrix}

\author{ Xiao-Jun Bi }
\affiliation{ Department of Physics, Tsinghua University, Beijing 100084, 
People's Republic of China}
\email[Email: ]{bixj@mail.tsinghua.edu.cn}

\author{ Yuan-Ben Dai }
\affiliation{ Institute of Theoretical Physics,
 Academia Sinica, P.O. Box 2735, Beijing 100080, People's Republic of China }
\email[Email: ]{dyb@itp.ac.cn}

\date{\today}

\begin{abstract}
The tiny neutrino masses measured in the neutrino oscillation experiments
can be naturally explained by the supersymmetric see-saw mechanism. 
If the supersymmetry breaking is mediated by gravity, the see-saw models may
predict observable lepton flavor violating effects. 
In this work, we investigate the lepton flavor violating process $\mu\to e\gamma$
in the kind of neutrino mass models based on the 
idea of the ``lopsided'' form of the charged lepton mass matrix.
The constraints set by the muon anomalous magnetic
moment are taken into account. We find 
the present models generally predict a much larger 
branching ratio of $\mu\to e\gamma$ than the experimental limit.
Conversely, this process may give strong constraint on the lepton flavor 
structure. Following this constraint we then find
a new kind of the charged lepton mass matrix. The feature of the
structure is that both the elements between the $2-3$ and $1-3$ generations 
are ``lopsided''. This structure produces a very small $1-3$ mixing 
and a large $1-2$ mixing in the charged lepton sector, which naturally 
leads to small $Br(\mu\to e\gamma)$ and the LMA solution for the solar neutrino
problem.
\end{abstract}

%\pacs{}
\preprint{TUHEP-TH-02135}

\maketitle

\section {introduction}

There are two remarkable features
of the neutrino parameters measured in the atmospheric and solar neutrino 
fluxes experiments, \textit{i.e.}, the extreme smallness of the neutrino
masses and the large neutrino mixing angles. The maximal atmospheric neutrino
mixing and the large mixing angle MSW solution of the solar neutrinos (LMA) are
most favored by the present experimental data\cite{lma}. The two features show 
that neutrinos have quite different properties compared with quarks and charged
leptons.  It is hoped that 
understanding the neutrino flavor structure may give important feedback
on, or even be the key to solve, the flavor problem, which includes 
how to understand the observed quark and charged lepton spectrum and mixing. 

Great enthusiasm has been stimulated on studying the neutrino physics and 
hundreds of models have been built and published 
in the recent years\cite{barr}. All the
authors tried to reflect the two notable features in their neutrino mass models.
The smallness of the neutrino masses is quite well understood now. See-saw 
mechanism seems the most natural and economic way to present tiny neutrino 
masses.  It is interesting to notice that the study of 
neutrino physics may provide us some information at very high energy scale
in the see-saw models. 

As for the large mixing angles, it is fare to say that no satisfying and
generally accepted
explanation is found until now; still, much progress has been achieved in
this aspect.  As we know that the neutrino mixing is actually the mismatch 
between the bases of the charged leptons and the active left-handed (LH) 
neutrinos, the neutrino models depending on the see-saw mechanism 
 are generally divided into 
two classes according to the origin of the large mixing angles coming from, 
\textit{i.e.}, those from the right-handed (RH) neutrino Majorana
mass matrix $M_R$, and transferring to the active LH neutrinos through the 
see-saw mechanism, 
\be
\label{ss}
M_\nu=-M_NM_R^{-1}M_N^T\ \ , 
\ene
where $M_N$ is the neutrino Dirac mass 
matrix, and those from the charged lepton mass matrix $M_L$.

See-saw mechanism is usually achieved in a supersymmetric grand unified 
model, which has the advantage of gauge couplings unification and avoidance of
the standard model hierarchy problem as well. 
Since leptons and quarks are placed in the same multiplets in grand unified
models and their mass matrices are related together, the question 
``How to keep small quark mixing while having large 
neutrino mixing'' is raised. The first class of models %mentioned above 
can naturally keep small quark mixing because
large neutrino mixing is from the RH Majorana neutrino mass matrix $M_R$, 
which is unrelated to the quark sector.  Several 
models are built based on this idea by different choice of the form of $M_R$
and give satisfied predictions\cite{first}. 
As for the second class of neutrino models, to avoid large quark
mixing, $M_L$ is usually chosen very asymmetric, or ``lopsided''. Then the 
large LH lepton mixing is related to the large RH quark
mixing, which is unobservable. A large number of
models have been built based on this elegant idea\cite{albright,order,lopsid}. 
%It seems that ``lopsided'' form of the charged lepton 
%mass matrix is the only known general guideline to give large neutrino
%mixing angles while keeping small quark mixing angles in neutrino mass
%model building. 

Neutrino oscillation means the neutrino flavor number is broken, which is
conserved in the standard model (SM) as an ``accidental'' global symmetry.
Lepton flavor number is another conserved quantum number as a global symmetry 
in the SM.
However, it is believed that the SM is only a low energy effective theory
and it must be extended to a fundamental theory at the high energy scale. 
These global symmetries are then broken in most extensions of the SM. 
Especially the
non-zero neutrino masses will also break the lepton flavor symmetry and lead to
lepton flavor violating (LFV) processes, such as 
$\mu\to e\gamma$, $\mu-e$ conversion,
$\mu\to eee$, $\tau\to\mu\gamma$, $\tau\to e\gamma$, $Z\to \mu\tau$ and so on. 
On one hand, the LFV processes are closely related to the neutrino oscillations.
On the other hand, they provide different information of the
lepton flavor structure. 
Combining study of the two kinds of processes may give us more comprehensive
understanding of the lepton flavor structure, even leading us to 
distinguish the origin of the large neutrino mixing angles.
At least, the LFV processes may give strong constraints on the 
neutrino mass models since they are measured quite accurately and the sensitivity
will be increased by three or more orders of magnitude in the near future\cite{mueg,exp,nexp}.

Recently there appear several works on the topic of lepton flavor violation
based on supersymmetric see-saw models\cite{casas,baek,carv,sato}. These works either give predictions of 
the LFV processes based on a specific neutrino mass model 
or give a general analysis based on the neutrino oscillation experimental data 
and the assumptions of the form of the RH neutrino mass matrix. 
For example, Ref. \cite{casas} discusses the branching ratio of $\mu\to e\gamma$
when the RH neutrinos are degenerate or completely hierarchical so that only
the heaviest eigenvalue is relevant to the form of $M_\nu$. Under these 
assumptions the LFV effects are directly related to the neutrino 
oscillation parameters. Actually many viable models do not belong to
these kinds. A general discussion of LFV for all neutrino models may be very
difficult. In our work, we focus on the LFV effects of
the kind of neutrino mass models exist in the literature based on the idea of
``lopsided'' form of the charged lepton mass matrices.
In this kind of models where the large $\nu_\mu-\nu_\tau$ mixing in neutrino oscillation is actually 
due to the mixing in the mass matrix of the charged leptons, instead of 
the neutrinos, it is thus natural to expect large LFV effects. In one of our
previous works we calculated the branching ratios of $\tau\rightarrow\mu\gamma$
and $Z\rightarrow\tau\mu$ generally in the ``lopsided'' models, 
independent of the model details and concluded that the process 
$\tau\rightarrow\mu\gamma$ might be detected in the next generation 
experiments\cite{bi}.  We will show in this work that the branching ratio of 
$\mu\rightarrow e\gamma$ is already larger than the 
present experimental limit in typical models of this kind proposed in the literature.
We then take $Br(\mu\rightarrow e\gamma)$ as a constraint which must be satisfied
by the models and find an interesting structure of the charged lepton mass matrix.
The notable feature of this model is that in addition to the
large $2-3$ element of the matrix the $1-3$ element is of order 1 too. 
This structure can suppress the branching ratio of $\mu\rightarrow e\gamma$
and lead to the LMA solution of the solar neutrino problem at the same time. 

%The calculation is performed within the framework of supersymmetry.
Since there is strong correlation between the muon radiative decay and the
muon anomalous magnetic dipole moment ($g_\mu-2$), we furthermore consider
the constraints on the SUSY parameter space set by the recent BNL E821 
experiment\cite{bnl}. After the correction to the sign of the light-by-light
term in the theoretical calculations\cite{musign}, there is only a 1.6$\sigma$
discrepancy between the measured value of the muon anomalous magnetic dipole
moment and the value predicted in the SM\cite{ng2}.
%The discrepancy, if finally confirmed, indicates new physics at the TeV scale
%in the lepton sector.

The paper is organized as follows. The Sec. II explains how the LFV effects 
are produced in the supersymmetric see-saw models and presents some analytic 
results of our calculation.
In Sec III we explain the basic idea of ``lopsided'' models and show the 
numerical results of $Br(\mu\to e\gamma)$ predicted by the models.
%why the ``lopsided'' models can give a general
%prediction of the branching ratio of $\mu\rightarrow e\gamma$.
%We briefly introduce our procudure of numerically
%solving the renormalization group equations (RGEs) in Sec IV and
We then present our new model in Sec IV and its predictions of $Br(\mu\to e\gamma)$.
Finally we give a conclusion of our work and discuss the implications
of the future experimental results of LFV processes in Sec V.

\section{some analytic results}

\subsection{Origin of LFV and related formulas}

In the pure SM the lepton flavor is strictly conserved.
If the SM is extended with massive and non-degenerate neutrinos, as suggested
by the atmospheric and solar neutrino fluxes experiments,
the LFV processes may be induced, in analogy to the Kobayashi-Maskawa (KM)
mechanism.  However, such processes are highly suppressed due to
the smallness of the neutrino masses. The branching ratio is proportional to
${\delta M^2_\nu}/{M^2_W}$ which is hopeless to be observed \cite{early}.

When supersymmetry enters the theory the scene changes completely. The LFV 
may also be induced
through the generation mixing of the soft breaking terms in the
lepton sector, \textit{i.e.}, the off-diagonal terms of the
slepton mass matrices $(m^2_{\tilde{L}})_{ij}$, $(m^2_{\tilde{R}})_{ij}$
 and the trilinear coupling $A^L_{ij}$. However, the 
present experimental bounds on the LFV processes give
very strong constraints on these off-diagonal terms, 
with the strongest constraint coming from $BR(\mu\to e\gamma)$
($<1.2\times 10^{-11}$)\cite{mueg}.

A generally adopted way to 
avoid these dangerous off-diagonal terms is to impose the 
universality constraints on the soft terms at the SUSY breaking scale,
such as in the gravity-mediated\cite{sgra} or gauge-mediated\cite{gmsb}
SUSY breaking scenarios. The universality assumption is certainly the most
conservative supposition we can make when we analyze the LFV effects.
Under the universality assumption off-diagonal terms can be
induced through quantum effects by the lepton flavor changing operators
existing at high energy scale, which 
are necessary to produce the neutrino mixing. 

We calculate these off-diagonal soft terms in the see-saw models, starting 
from the universal initial values at the GUT scale, by numerically solving the 
renormalization group equations (RGEs).% for these soft terms.
Then we calculate the branching ratio of $\mu \to e \gamma$ induced by these
terms. 

In a supersymmetric see-saw model, the RH neutrinos are active
at the energy scale above $M_R$(We will use the same symbol 
as the RH Majorana mass matrix $M_R$ to represent the energy scale).
The superpotential of the lepton sector is then given by
\be
W=Y_N^{ij} \hat{H}_2\hat{L}_i\hat{N}_j+Y_L^{ij}\hat{H}_1\hat{L}_i\hat{E}_j
+\frac{1}{2}M_R^{ij}\hat{N}_i\hat{N}_j+\mu \hat{H}_1\hat{H}_2\ \ ,
\ene
where $Y_N$ and $Y_L$ are the neutrino and charged lepton Yukawa coupling matrices respectively, 
$i$, $j$ are the generation indices. 
%Anti-symmetric tensor $\epsilon^{ab}$ is implict to contract the SU(2) doublets with $\epsilon^{12}=-1$.
In general, $Y_N$ and $Y_L$
can not be diagonalized simultaneously and lead to LFV interactions.
The three matrices $Y_N$, $Y_L$ and $M_R$ can be diagonalized by
\bea
\label{ul}
Y^\delta_L&=&U_L^\dagger Y_L U_R\ \ ,\\
\label{vl}
Y^\delta_N&=&V_L^\dagger Y_N V_R\ \ ,\\
M_R^\delta&=&X^T M_R X\ \ ,
\enea
respectively, where $U_{L,R}$\ , $V_{L,R}$ and $X$ are unitary matrices. The RH
neutrinos can be easily integrated out one by one on the bases where $M_R$ is
diagonal and then the relation~(\ref{ss}) is recovered. Define
\be \label{vd} V_D=U_L^\dagger V_L\ , \ene
which is analog to the KM matrix $V_{KM}$ in the quark sector.
Then $V_D$ determines the lepton flavor mixing. We can see that $V_D$ is 
determined by the LH mixing of the Yukawa coupling matrices $Y_L$ and $Y_N$
and only exists above the scale $M_R$. The MNS mixing matrix,
which describes the low energy neutrino mixing, is defined by
\be
V_{MNS}=U_L^\dagger U_\nu,
\ene
where $U_\nu$ is the unitary matrix diagonalizing the neutrino mass matrix 
$M_\nu$ in Eq.~(\ref{ss}). If $V_{MNS}$ and $V_D$ are both determined
by the neutrino oscillation and LFV experiments respectively,
it is possible to infer whether the large neutrino mixing is coming from the
charged lepton sector or not. We can furthermore learn some information about the 
structure of the RH neutrino mass matrix.

The corresponding soft breaking terms for the lepton sector are
\cite{rosiek}
\bea
{\cal{L}}_{soft}&=&-m_{H_1}^2H_1^\dagger H_1-m_{H_2}^2H_2^\dagger
H_2
-(m_{\tilde{L}}^2)^{ij}\tilde{L}_i^\dagger\tilde{L}_j
-(m_{\tilde{E}}^2)^{ij}\tilde{E}_i^*\tilde{E}_j-(m_{\tilde{N}}^2)^{ij}
\tilde{N}_i^*\tilde{N}_j\nonumber\\
&&+\left(B\mu H_1H_2+\frac{1}{2}BM_R^{ij}\tilde{N}_i^*\tilde{N}_j^*
+(A^LY_L)^{ij}H_1\tilde{L}_i\tilde{E}_j\right.\nonumber\\
&&\left.+(A^N Y_N)^{ij}H_2\tilde{L}_i\tilde{N}_j+h.c.\right)\ \ .
\enea
We assume the universal condition at the GUT scale
\bea
m_{H_1}^2&=&m_{H_1}^2=m_0^2\ \ ,\\
m_{\tilde{L}}^2&=&m_{\tilde{R}}^2=m_{\tilde{N}}^2=m_0^2\ \ ,\\
A^L&=&A^N=A_0\ \ .
\enea

The above LFV effects in the superpotential, determined by $V_D$,
then transfer to the soft terms through quantum effects and induce non-diagonal
terms from the initial universal form.
This is clearly shown by the following 
%The dominant contribution of low energy lepton flavor violation comes from
%the generation mixings of $m_{\tilde{L}}^2$. The 
RGE for $m_{\tilde{L}}^2$, which gives the dominant contribution to the low
energy LFV processes, %such as $\mu\to e\gamma$ and so on, 
\bea
\mu\frac{dm_{\tilde{L}}^2}{d\mu}&=&\frac{2}{16\pi^2}\left[-\Sigma c_ig_i^2M_i^2+
\frac{1}{2}[Y_NY_N^\dagger m_{\tilde{L}}^2+m_{\tilde{L}}^2Y_NY_N^\dagger]+
\frac{1}{2}[Y_LY_L^\dagger m_{\tilde{L}}^2+m_{\tilde{L}}^2Y_LY_L^\dagger]\right.\nonumber\\
&&\left.+Y_Lm_{\tilde{E}}^2Y_L^\dagger+m_{H_D}^2Y_LY_L^\dagger+E_AE_A^\dagger+
Y_N m_{\tilde{N}}^2Y_N^\dagger+m_{H_U}^2Y_NY_N^\dagger+N_AN_A^\dagger\right]
\enea
with $E_A=A^L\cdot Y_L$, $N_A=A^N\cdot Y_N$ and $g_i$, $M_i$ being the
gauge coupling constants and gaugino masses respectively.

In the basis where $Y_L$ and $M_R$ are diagonal, an approximate formula for
the off-diagonal terms of $m_{\tilde{L}}^2$ applicable for $\delta m^2\ll m_0^2$ is given by
\bea
\left(\delta m_{\tilde{L}}^2\right)_{ij}&\approx &\frac{1}{8\pi^2}(Y_NY_N^\dagger)_{ij} (3+a^2)m_0^2\log
\frac{M_{GUT}}{({M_R})_3}+\cdots\nonumber\\
&\approx&\frac{1}{8\pi^2}(V_D)_{i3}(V_D^*)_{j3}\cdot Y_{N_3}^2(3+a^2)m_0^2\log\frac{M_{GUT}}{({M_R})_3}+\cdots \ \ ,
\label{dm}
\enea
where in the diagonalized Yukawa matrix $Y_N^\delta$ only the (3,3) element
$Y_{N_3}$ is kept under the assumption of a hierarchical form of $Y_N$.
We will use the GUT-motivated assumption 
%that $Y_N$ has a similar hierarchical structure 
%with that of the up quarks, which leads to 
$Y_{N_3}\approx Y_t$ as an order estimate. 
The `$a$' is the universal trilinear coupling given by $A_0=a m_0$.
$({M_R})_3$ is the mass of the heaviest RH neutrino. The ellipsis dots 
`$\cdots$' in Eq. (\ref{dm}) represent the corrections due to the RGE running
below the scale $({M_R})_3$, where the heaviest RH neutrino is decoupled.
% and $V_R^\dagger X$ is no longer unitary. 
%the nondominant LFV effect, which depends on $V_R^\dagger X$. If 
%In the case of $V_R^\dagger X\approx 1$, we have
%\be
%\left(\delta m_{\tilde{L}}^2\right)_{ij}\approx 
%\sum_k\frac{1}{8\pi^2}(V_D)_{ik}(V_D^*)_{jk}\cdot Y_{N_k}^2(3+a^2)m_0^2\log\frac{M_{GUT}}{({M_R})_k}\ \ .
%\ene
If there is large mixing between the first two generations and the third 
generation in $V_R^\dagger X$, for example, the large $2-3$ mixing, 
we will have another
two terms besides the term in Eq.~(\ref{dm}), \textit{i.e.},
\be
%\cdots\approx 
\frac{1}{8\pi^2}(V_D)_{i3}(V_D^*)_{j3}\cdot Y_{N_3}^2(3+a^2)m_0^2
\left((1-|(V_R^\dagger X)_{33}|^2)\log\frac{({M_R})_3}{({M_R})_2}+ 
|(V_R^\dagger X)_{31}|^2\log\frac{({M_R})_2}{({M_R})_1}\right).
\ene
The above term is negligible in the case of $V_R^\dagger X\approx 1$.
%This formula is very convenient for an quanlitative analysis and we will use it 
%frequently.

\subsection{ Some analytic formulas }

In this subsection we give the analytic expressions for the branching ratio of
the LFV process $\mu\to e\gamma$ and the muon anomalous magnetic dipole moment
 within the supersymmetric framework. 

\begin{center}
\begin{figure}
\begin{picture}(500,250)(20,50)
\ArrowLine(20,200)(65,200)
\ArrowLine(65,200)(185,200)
\ArrowLine(185,200)(230,200)
\ArrowLine(270,200)(315,200)
\ArrowLine(315,200)(435,200)
\ArrowLine(435,200)(480,200)
\DashArrowArcn(125,200)(60,180,0){4}
\DashArrowArcn(375,200)(60,180,0){4}
\Photon(125,185)(180,130){4}{7}
\Text(150,135)[]{\huge $\gamma$}
\Text(170,250)[l]{\huge $\tilde{\nu}_\alpha$}
\Text(30,185)[]{\huge $\mu$}
\Text(220,185)[]{\huge $e$}
\Text(125,215)[]{\Large $\chi^-_a$}
\Text(40,215)[]{\huge $p_\mu$}
\Text(210,215)[]{\huge $p_e$}
\Text(195,125)[]{\huge $q$}
 
\Photon(380,270)(450,305){4}{7}
\Text(440,285)[]{\huge $\gamma$}
\Text(328,250)[r]{\huge $\tilde{l}_\alpha$}
\Text(280,185)[]{\huge $\mu$}
\Text(470,185)[]{\huge $e$}
\Text(375,185)[]{\Large $\chi^0_a$}
\end{picture}
 
\vspace*{-1.5cm}
\caption{\label{fig1} Feynman diagrams for the process $\mu \to e\gamma$
by exchanging charginos and neutralinos respectively.}
\end{figure}
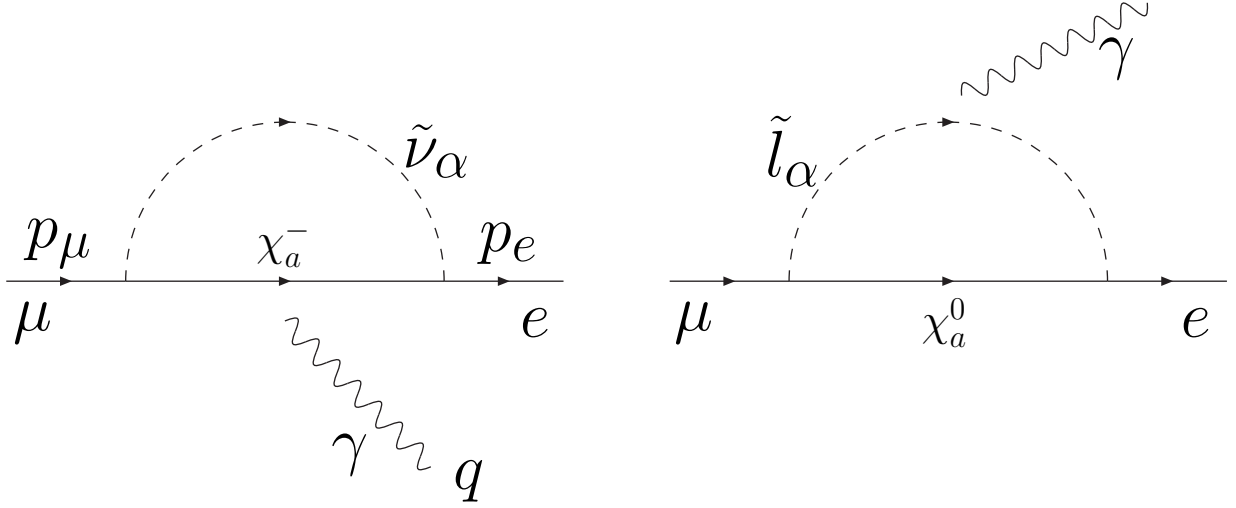
\end{center}
The LFV decay of muon occurs
through photon-penguin diagrams including sleptons, neutralinos and charginos,
shown in FIG. \ref{fig1}. The amplitude of the 
charged lepton radiative decay can be 
written in a general form
\be
\label{amp}
M=em_i \bar{u}_j(p_j)i\sigma_{\mu\nu}q^\nu
(A^{ij}_LP_L+A^{ij}_RP_R)u_i(p_i)\epsilon^\mu (q)\ \ ,
\ene
where $P_{L,R}=\frac{1}{2}(1\mp \gamma_5)$ are the chirality projection operators.
The $i$, $j$ represent initial and final lepton flavors respectively.
The most convenient way to calculate $A_L$ and $A_R$ is to pick up the one
loop momentum integral contributions which are proportional to
$\bar{u}_j(p_j)P_{L,R}u_i(p_i)2p_i\cdot\epsilon$ respectively.
The neutralino exchanging contribution is
\bea
A_L^{(n)}&=&-\frac{1}{32\pi^2}(\frac{e}{\sqrt{2}\cos\theta_W})^2\frac{1}
{m_{\tilde{l}_\alpha}^2}\left[{B^{j\alpha a}}^*B^{i\alpha a}F_1(k_{\alpha a}) 
+\frac{m_{\chi_a^0}}{m_i}{B^{j\alpha a}}^*
A^{i\alpha a}F_2(k_{\alpha a})\right] \ ,\\
A_R^{(n)}&=&A_L^{(n)}\ (B\leftrightarrow A)\ ,
\enea
where 
\bea
F_1(k)&=&\frac{1-6k+3k^2+2k^3-6k^2\log k}{6(1-k)^4}\ \ ,\\
F_2(k)&=&\frac{1-k^2+2k\log k}{(1-k)^3}\ \ ,
\enea
with $k_{\alpha a}={m_{\chi_a^0}^2}/{m_{\tilde{l}_\alpha}^2}$. $A$ and $B$ are
the lepton--slepton--neutralino coupling vertices given by
\bea
A^{i\alpha a}&=&\left(Z_{\tilde{L}}^{i\alpha}(Z_N^{1a}+Z_N^{2a}\cot\theta_W)-\cot\theta_W\frac{m_{l_i}}{
M_W\cos\beta}Z_{\tilde{L}}^{(i+3)\alpha}{Z_N^{3a}}\right)\ \ ,\\
B^{i\alpha a}&=&-\left(2Z_{\tilde{L}}^{(i+3)\alpha}{Z_N^{1a}}^*+\cot\theta_W
\frac{m_{l_i}}{M_W\cos\beta}Z_{\tilde{L}}^{i\alpha}{Z_N^{3a}}^*\right)\ \ ,
\enea
with $Z_{\tilde{L}}$ being the $6\times 6$ slepton mixing matrix and $Z_N$ 
being the neutralino mixing matrix given in Ref. \cite{rosiek,bi}. 
The corresponding contribution coming from exchanging charginos is
\bea
A_L^{(c)}&=&\frac{g_2^2}{32\pi^2}{Z_{\tilde{\nu}}^{i\alpha}}^*Z_{\tilde{\nu}}^{j\alpha}
\frac{1}{m_{\tilde{\nu}_\alpha}^2}\left[ Z^-_{2a}{Z^-_{2a}}^*
\frac{m_im_j}{2M_W^2\cos^2\beta}
F_3(k_{\alpha a})\right. \nonumber \\
&&\left. +\frac{m_{\chi^-_a}}{\sqrt{2}M_W\cos\beta}Z^+_{1a}Z^-_{2a}
\frac{m_j}{m_i}
F_4(k_{\alpha a})\right]\ \ ,\\
\label{imp}
A_R^{(c)}&=&\frac{g_2^2}{32\pi^2}{Z_{\tilde{\nu}}^{i\alpha}}^*Z_{\tilde{\nu}}^{j\alpha}
\frac{1}{m_{\tilde{\nu}_\alpha}^2}\left[ Z^+_{1a}{Z^+_{1a}}^*F_3(k_{\alpha a})
+\frac{m_{\chi^-_a}}{\sqrt{2}M_W\cos\beta}{Z^+_{1a}}^*{Z^-_{2a}}^*
F_4(k_{\alpha a})\right]\ \ ,
\enea
where 
\bea
F_3(k)&=&\frac{2+3k-6k^2+k^3+6k\log k}{6(1-k)^4}\ \ ,\\
F_4(k)&=&\frac{3-4k+k^2+2\log k}{(1-k)^3}\ \ ,
\enea
with $k_{\alpha a}={m_{\chi_a^-}^2}/{m_{\tilde{\nu}_\alpha}^2}$. 
$Z_{\tilde{\nu}}$, $Z^+$ and $Z^-$ are the mixing matrices of sneutrinos and
charginos as given in Ref. \cite{rosiek,bi}.

The branching ratio of $\mu\to e\gamma$ is given by
\be
BR(\mu\to e\gamma)=\frac{\alpha_{em}}{4\pi}m_\mu^5(|A^{12}_L|^2+|A^{12}_R|^2)/
\Gamma_\mu \ \ ,
\ene
with $\Gamma_\mu=2.996\times 10^{-19}GeV$\cite{pdg}.

To identify the dominant contribution one may use the mass
insertion approximation\cite{hisano,casas}. Under this approximation and when
$\tan\beta$ is large we have
\be
\label{appbr}
Br(\mu\to e\gamma)\ \sim\ \frac{\alpha^3}{G_F^2}\frac{ [(m^2_{\tilde{L}})_{12}]^2 }{m_s^8}\tan^2\beta\ \ ,
\ene
where $m_s$ represents general slepton mass and ``$\sim$'' means some constants
are omitted on the right-hand side. From this expression we
can clearly see that the supersymmetric contribution to $Br(\mu\to e\gamma)$
is proportional to $\tan^2\beta$ and the first two generation slepton mass
mixing.
%We find the supersymmetric contributions come dominantly from 
%$A_R^{(c)}$~.

In the computation of the LFV branching ratios we considered the 
constraints on the supersymmetric parameter space set by the 
BNL E821 experiment of the muon anomalous magnetic moment\cite{bnl}. 
We give the related analytic formulas in the following. 

The amplitude for the photon-muon-muon coupling in the limit of the
photon momentum $q$ tending to zero can be written as
\be
M'=ie\bar{u}_\mu\left\{ \gamma^\alpha+a_\mu\frac{i\sigma^{\alpha\beta}q_\beta}{2m_\mu}\right\}u_\mu \epsilon_\alpha(q)\ \ ,
\ene
with $a_\mu$ being the muon anomalous magnetic moment.
The discrepancy is give by\cite{ng2}
\be
\delta a_\mu=a_\mu^{exp}-a_\mu^{SM}=26 (\pm 16)\times 10^{-10}\ \ .
\ene

The supersymmetric contribution to $\delta a_\mu$ comes from the same 
photon-penguin diagrams producing $\mu\to e\gamma$ decay by replacing the
final state by muon. The analytic expressions for $\delta a_\mu$ from the
neutralino and chargino exchanging contributions are 
\bea
\label{auneu}
\delta a_\mu^{(n)}&=&-\frac{1}{32\pi^2}\frac{e^2}{\cos^2\theta_W}\frac{m_\mu^2}{m_{\tilde{l}_\alpha}^2}\cdot \nonumber\\
&&\left[({A^{i\alpha a}}^{*}A^{i\alpha a}+{B^{i\alpha a}}^{*}B^{i\alpha a}) F_1(k_{\alpha a})+\frac{m_{\chi_a^0}}{m_\mu}
Re({A^{i\alpha a}}^*B^{i\alpha a})F_2(k_{\alpha a})\right]\ \ ,  \\
\text{and}&&\nonumber\\
\label{aucha}
\delta a_\mu^{(c)}&=&\frac{g_2^2}{16\pi^2}\frac{m_\mu^2}{m_{\tilde{\nu}_\alpha}^
2}Z_{\tilde{\nu}}^{i\alpha}{Z_{\tilde{\nu}}^{i\alpha}}^*\cdot \nonumber\\
&&\left[( {Z_{1a}^+}^*Z_{1a}^++\frac{m_\mu^2}{2M_W^2\cos\beta}{Z_{2a}^-}^*Z_{2a}^- )F_3(k_{\alpha a})+\frac{m_{\chi_a^-}}{\sqrt{2}M_W\cos\beta}Re(Z_{1a}^+Z_{2a}^-)
F_4(k_{\alpha a})\right] \ .
\enea 
respectively. 
%Under the mass insertion approximation the expressions are of 
%the form \cite{feng}
%\be
%\frac{g_i^2}{16\pi^2}m_\mu^2\mu M_i\tan\beta\ F\ ,
%\ene
%where $i=1,2$ and $F\propto M_{SUSY}^{-4}$.

Comparing the expressions for the amplitude of muon decay and muon anomalous 
magnetic moment we find a striking resemblance. 
%between the effective operators that generate $\mu\to e\gamma$ and $\delta a_\mu$ 
Because a mass insertion is necessary to give correct fermion chirality in 
Eq. (\ref{amp}), $|A_R|$ dominates over the contributions. At large $\tan\beta$ a direct relation
between the two quantities is then found under the mass insertion 
approximation\cite{carv}
\be
\label{rela}
Br(\mu\to e\gamma)\approx \frac{192 \pi^3\alpha}{G_F^2m_\mu^4}\times(\delta
a_\mu)^2\times\kappa^2\ \ ,
\ene
where $\kappa\equiv A_R^{12}/A_R^{22}$. It is easy to check that
Eq. (\ref{rela}) is consistent with Eq. (\ref{appbr}) and $\kappa
\approx ({\delta m_{\tilde{L}}^2})_{12}/m_s^2$ if we notice that
$Z_{\tilde{\nu}}^{i\alpha}({m_{\tilde{\nu}_\alpha}^2})^{-1}{Z_{\tilde{\nu}}^{j\alpha}}^*
\approx -({\delta m_{\tilde{L}}^2})_{ij}/m_s^4$ with $m_s$ being the common slepton mass.
%provided all the superparticles have approximate the same masses. 
Assuming $\delta a_\mu$ is due to the 
supersymmetric corrections, Eq. (\ref{rela}) may give
constraints on the numerical results of the LFV branching ratio
if the flavor mixing parameter $\kappa$ can be predicted in an explicit model.

\section{$\mu\to e\gamma$ in the ``lopsided'' neutrino mass models}

\subsection{ Basic idea of ``lopsided'' neutrino mass models }

In a SU(5) grand unified model, the LH charged leptons are in the same
multiplets as the CP conjugates of the RH down quarks. This feature leads to
the fact that the mass matrix
for the charged leptons is related closely in form to the transpose of the mass
matrix of the
down quarks. The basic idea of ``lopsided'' models is that the charged lepton
and the down quark mass matrices have the approximate forms as\cite{albright,order,lopsid}
\begin{equation}
\label{lop}
M_L\ \sim\ \left( \begin{array}{ccc}
0 & 0 & 0 \\ 0 & 0 & \sigma \\ 0 & \epsilon & 1 \end{array} \right) m_D,\ \
M_D\ \sim \ \left( \begin{array}{ccc}
0 & 0 & 0 \\ 0 & 0 & \epsilon \\ 0 & \sigma & 1 \end{array} \right) m_D
\end{equation}
with $\epsilon \ll 1$ while $\sigma \sim 1$ and zeros representing small
entries. In the case of the charged leptons
$\sigma$ controls the mixing of the second and the third families of the LH
leptons (Here we use the convention that the LH doublet multiplies the Yukawa
coupling matrix from the left side while the RH singlet from the right side),
which contributes to the atmospheric neutrino mixing and makes the
mixing angle $\theta_{atm}$ large, while $\epsilon$ controls the mixing of the
second and the third families of the RH leptons, which is not observable at
low energy. For the quarks the reverse is the case: the small
$\mathcal{O}(\epsilon)$ mixing
is in the LH sector, accounting for the smallness of $V_{cb}$, while the large
$\mathcal{O}(\sigma)$ mixing is in the RH sector, which is not observable.

The crucial element in the approach is the ``lopsided'' form of the mass
matrices for the charged leptons and the down quarks and
that $M_L$ being similar to the transpose of $M_D$. The relation $M_L\sim M_D^T$
can be achieved in grand unified models based on larger groups where SU(5) is a
subgroup and plays the critical role to give the relation. This relation
can also be achieved in models with abelian or non-abelian flavor symmetries,
no matter they base on grand unification or not. Generally there is no such
relationship between $M_N$ and the up quark
mass matrix $M_U$ in SU(5). They are usually assumed to be not lopsided and give 
small mixing in the literature. However, we shall use $({M_U})_{33}=({M_N})_{33}$ in our calculation, which is valid in the simplest SO(10) model.

As for the large
mixing between the first two generations in the neutrino oscillation, which is assumed to explain the
solar neutrino problem, it is usually given by the LH neutrino mass matrix
$M_\nu$ in the literature\cite{barr,albright,order,lopsid}. Thus the mixing angle 
$\theta_{12}$ of the LH charged leptons is usually small. 

\subsection{Order estimate of $\mu\to e\gamma$ in the ``lopsided'' models}

We first give an order estimate of the branching 
ratio of $\mu\to e\gamma$ in the ``lopsided'' models. 
The quantitative results are given by
solving the RGEs numerically and presented in the next subsection. 

The starting point is Eq. (\ref{dm}). In the ``lopsided'' models
there can be no large mixing in $V_R^\dagger X$ and the RGEs running
below $({M_R})_3$
in Eq. (\ref{dm}) is ignored.
Under the assumption ${Y_N}_3\approx Y_t$,
%$M_R$ is determined by the requirement of giving correct neutrino mass by
%the see-saw mechanism. 
the ratio between the off-diagonal term $(\delta
m_{\tilde{L}}^2)_{12}$ and the diagonal term, which is approximately $m_0$,
is determined by ${(V_D)}_{13}$, ${(V_D)}_{23}$ and $({M_R})_3$.

$V_D$ is defined in Eq. (\ref{vd}). We estimate $V_D$ through an analysis 
similar to the analysis in Ref. \cite{barr} in which the order of
the MNS matrix element $U_{e3}$ is determined.  Writing $U_L$ in the form
\begin{equation}
\label{angles}
U_L = \left( \begin{array}{ccc}
1 & 0 & 0 \\ 0 & \overline{c}_{23} & \overline{s}_{23} \\
0 & - \overline{s}_{23} & \overline{c}_{23} \end{array} \right)
\left( \begin{array}{ccc} \overline{c}_{13} & 0 & \overline{s}_{13} \\
0 & 1 & 0 \\ - \overline{s}_{13} & 0 & \overline{c}_{13} \end{array}
\right) \left( \begin{array}{ccc} \overline{c}_{12} & \overline{s}_{12} &
0 \\ - \overline{s}_{12} & \overline{c}_{12} & 0 \\ 0 & 0 & 1
\end{array} \right),
\end{equation}
where $\overline{s}_{ij} \equiv \sin \overline{\theta}_{ij}$, 
$\overline{c}_{ij} \equiv \cos \overline{\theta}_{ij}$
and $V_L$ in a similar way with the corresponding
angles being denoted by $\tilde{\theta}_{ij}$,
we get the expressions for $({V_D})_{13}$ and $({V_D})_{23}$,
\bea
\label{acvd13}
({V_D})_{13} & = & - \overline{s}_{12} s_{23} \tilde{c}_{13}
- \overline{s}_{13} \overline{c}_{12} c_{23} \tilde{c}_{13}
+ \overline{c}_{13} \overline{c}_{12} \tilde{s}_{13} \ \ , \\
({V_D})_{23} & = & \overline{c}_{12} s_{23} \tilde{c}_{13} +
\overline{c}_{13} \overline{s}_{12} \tilde{s}_{13}
- \overline{s}_{13} \overline{s}_{12} c_{23} \tilde{c}_{13}\ \ ,
\enea
with $s_{23}\equiv \sin(\tilde{\theta}_{23}-\bar{\theta}_{23})$.
Under the GUT-based assumption that $Y_N$ has a similar hierarchical 
structure with that of up quark, 
we expect that $\tilde{s}_{13}$ is of the same order as the corresponding
mixing angle, $\theta_U^{13}$, in the up quark mixing matrix. 
If we further assume that no accidental cancellation exists 
between the up and down quark mixing matrices, we then get 
$\theta_U^{13}\lesssim V_{td}\sim 0.008$\cite{pdg},
where $V_{td}$ is the 31 element of the CKM matrix. 
So, $\tilde{s}_{13}\lesssim 0.008$.
Similarly, we have $\tilde{s}_{23}\lesssim V_{ts}\cong 0.04$. 
One typically finds that 
$\bar{s}_{12}\sim \sqrt{m_e/m_\mu}\cong 0.07$ and $\bar{s}_{13}\approx
m_\mu/m_\tau \bar{s}_{12}\ll \bar{s}_{12}$\cite{barr,frit}. These two relations
hold in most viable models, where $m_\tau$ and $m_\mu$ get their masses from the 2-3 block of the mass matrix.
% except in the flavor democracy ones\cite{fdemo}.
Considering that $s_{23}\sim c_{23}\sim \mathcal O(1)$ in the ``lopsided'' 
models then we get 
\bea
\label{vd13}
({V_D})_{13} & \approx & - \overline{s}_{12} s_{23} 
\approx \bar{s}_{12} \bar{s}_{23} \approx 0.05\ \ ,\\
\label{vd23}
({V_D})_{23} & \approx &  s_{23} \approx -\bar{s}_{23} \approx -0.71\ \ .
\enea
The $({V_D})_{13}$ and $({V_D})_{23}$ are actually determined by $U_L$ 
solely in the ``lopsided'' models. This conclusion certainly depends on the 
assumption of the form of $Y_L$ and $Y_N$. However, it is actually correct 
in most published ``lopsided'' models\cite{albright,order}. 
We can actually relax this assumption, unless strong
cancellation taking place in Eq. (\ref{acvd13}), we always get that $({V_D})_{13}$ is 
$\mathcal{O}(0.05)$ or larger. 

$M_R$ is determined by using the see-saw relation Eq. (\ref{ss}) conversely
\be
M_R=- M_N^T M_\nu^{-1} M_N .
\ene
If we assume that $M_N$ has similar spectrum to the up quarks
and $M_\nu$ gives the large mixing for neutrino oscillation,
\textit{i.e.}, $V_{L,R}\approx U_{L,R}\approx I$ and $U_\nu\approx V_{MNS}$, 
we find the scale of $M_R$ is actually determined by the lightest neutrino mass,
with $({M_R})_3\gtrapprox 2.5\times 10^{15} GeV$. If $Y_L$ is lopsided
and $M_\nu$ is approximately diagonal, \textit{i.e.}, $U_L^\dagger\approx V_{MNS}$
and $U_\nu\approx I$, the scale of $M_R$ is then determined
by the heaviest neutrino mass, with $({M_R})_3\approx 4\times 10^{14} GeV$. 

Having known the values of the elements $(V_D)_{13}$, $(V_D)_{23}$ and $M_R$, 
we can give the order estimation of $\mu\to e\gamma$ by a simple calculation.
%By Eq. (\ref{rela}), we have $Br(\mu\to e\gamma)\sim 10^{-9} - 10^{-7}$, 
%which is $2\sim 4$ orders larger than the present experimental limit,
%corresponding to take $\delta a_\mu$ the $\mp 1.5\sigma$ values around the 
%experimental central value\footnote{The numerical result of Br($\mu\to e\gamma$)
%certainly does not depend on the experimental data of $\delta a_\mu$ directly.}.
At the large $\tan\beta$ case, the most important contribution is
coming from the second term in Eq. (\ref{imp}). Assuming that all the SUSY particles 
are of the scale $m_s$ we then have 
\bea
Br(\mu\to e\gamma)& \approx & \frac{\alpha_{em}}{4\pi} m_\mu^5
\left( \frac{g_2^2}{32\pi^2} \right)^2
\left( \frac{ ({m^2_{\tilde{L}}})_{12} }{m_s^4} \right)^2 
\left| \frac{ m_{\chi_a^-}Z_{1a}^+Z_{2a}^-F_4(k_a) }{\sqrt{2}M_W} \right|^2
\tan^2\beta /\Gamma_\mu \\
& \approx & C\cdot 10^{-7} 
\left( \frac{100 GeV}{m_s} \right)^4
\left( \frac{\tan\beta}{10} \right)^2\ ,
\enea
with $C$ being around $1\sim 10$.
This estimate means that unless all the SUSY particles tend to above $1 TeV$ and
$\tan\beta$ is not too large the 
branching ratio should be above the present experimental limit. 
The reason for such a large branching ratio is clear: the ``lopsided'' structure
enhances $({V_D})_{23}$ and $({V_D})_{13}$ at 
the same time, as shown in Eqs. (\ref{vd13}) and (\ref{vd23}).
The branching ratio is thus approximately proportional to $\bar{s}_{23}^4$,
which is about 2 orders larger if taking $\bar{s}_{23}\approx \sqrt{m_\mu/m_\tau}$
in the symmetric case.  Then the branching ratio may be at the same order of, 
or be slightly below, the present experimental limit.

\subsection{Numerical results}

In this subsection we give the numerical results of the branching
ratio of $\mu\to e \gamma$ in the ``lopsided'' models. When solving the RGEs we
%transform the RGE (\ref{RGE}) to the basis where $Y_L$ is diagonal and 
only keep the third generation Yukawa coupling 
eigenvalues $Y_{\tau}$ and $Y_{N_3}$ in $Y_L$ and $Y_N$ and ignore the contributions from the first two generations. We take
$({V_D})_{23}=-\bar{s}_{23}=-1/\sqrt{2}$ and
$({V_D})_{13}=\bar{s}_{12} \bar{s}_{23}=0.07/\sqrt{2}$ as the
typical values in this kind of models. We also show the results
when $\bar{s}_{12}$ is as small as $0.01$. Below $M_R$ we solve
the RGEs of MSSM where $Y_N$ is absent. The details about solving
the RGES are given in Ref. \cite{bi}. We present the dependence of
the results on the SUGRA parameters: $m_0$, $m_{1/2}$, $A_0$,
$\tan\beta$ and the sign of $\mu$ parameter.

We first constrain the SUGRA parameter space by the following conditions:
(1) electroweak spontaneous symmetry breaking is produced by radiative
corrections and correct vacuum expectation value is given; (2) the LSP of
the model be the lightest neutralino with its mass limit $m_{\chi^0_1} >
45 GeV$; (3) the lightest chargino is heavier than $94 GeV$ and
(4) the lightest charged slepton is heavier than $81 GeV$. $\delta a_\mu$
gives further constraints on the parameter space. It
always constrains the $\mu$ parameter to be positive.  We fix $A_0=m_0$
throughout the calculation, which does not affect the result much, and
take $M_R=4.5\times 10^{14} GeV$.
%The chargino mass limit gives a strong constraints on the parameter $m_{1/2}$.

%We keep the third generation Yukawa couplings in $Y_L$ and $Y_N$ in the RGEs 
%up to the second order of
%$Y_{N_3}$, \textit{i.e.}, the terms like $\delta m^2\cdot Y_{N_3}$ are kept with 
%$\delta m^2 \propto Y_{N_3}$ being the quantum correction. First order of $Y_\tau$
%is kept in the RGEs. The terms propotional to $Y_{N_{2,1}}$ and $Y_{\mu,e}$
%are ignored. 
%and ignore the others. 
%Since $Y_{N_3}$ is the most important element to raise the
%off-diagonal terms in the slepton mass matrix, it is simple to solve the RGEs 
%on the basis where $Y_N$ is diagonal above $M_R$. 
%We then rotate to the basis where $Y_L$ 
%is diagonal at the scale $M_R$. Below $M_R$ we solve the RGEs of MSSM. 
%we have ignored all the other contributions below this scale. 
%The details about solving the REGS are give in Ref. \cite{bi}.

\begin{figure}
\includegraphics[scale=0.6]{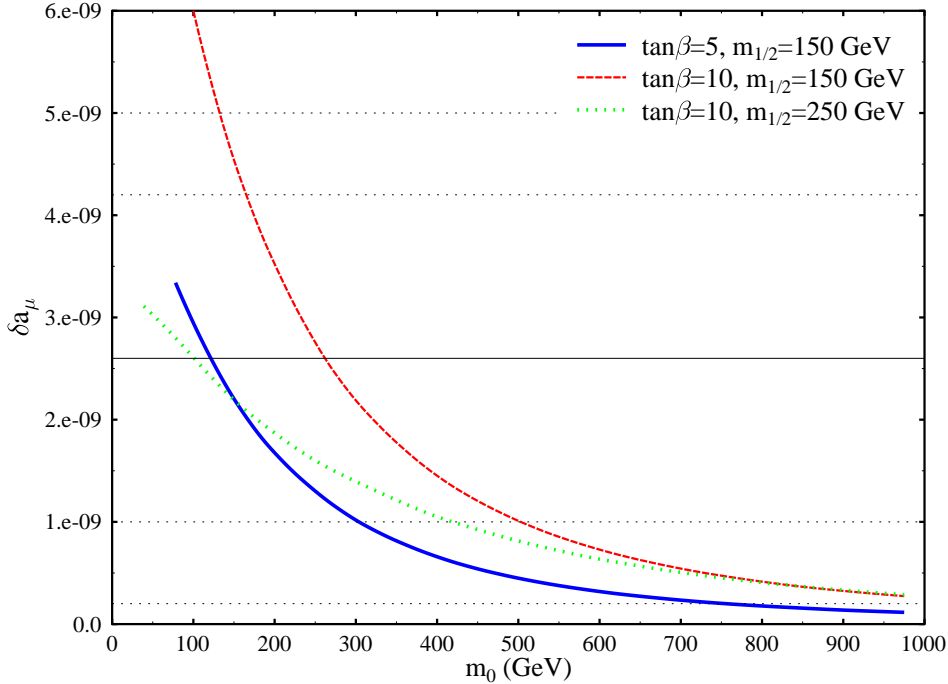}
\caption{\label{fig2} $\delta a_\mu$ as a function of $m_0$ for
$\tan\beta=5$, $m_{1/2}=150
GeV$ and for $\tan\beta=10$, $m_{1/2}=150, 250 GeV$ respectively. $A_0=m_0$ and
$\mu > 0$ are assumed. For $\mu < 0$, $\delta a_\mu$ is negative with almost
the same magnitude as the present corresponding values. The horizontal lines
represent the E821 central value and it $\pm 1\sigma$ and $\pm 1.5\sigma$ bounds. }
\end{figure}

In FIG. \ref{fig2} we plot $\delta a_\mu$ as a function of $m_0$ for $\tan\beta
=5, 10$ and $m_{1/2}=150, 250 GeV$ respectively.  The horizontal lines are
the E821 central value and its $\pm 1\sigma$ and $\pm 1.5\sigma$ bounds.
$m_{1/2}=150 GeV$ is almost the smallest value we can take considering
the chargino mass experimental limit. We find $\mu < 0$ always predicts
a $\delta a_\mu < 0$ with almost the same magnitude as the 
corresponding predictions for positive $\mu$. 
%$\delta a_\mu$ behaves just as what we expect: decreases quickly
%when $m_0$ and $m_{1/2}$ become large and increases as $\tan\beta$ becomes large.

The formulas for $\delta a_\mu$ we adopted, Eqs. (\ref{auneu}) and
(\ref{aucha}), are different
from those given by most authors because we use the full slepton and sneutrino
mass matrices including the off-diagonal terms. The numerical results,
however, do not depend on the mixing angles, just as expected.
This can be regarded as a check for our program.

\begin{figure}
\includegraphics[scale=0.8]{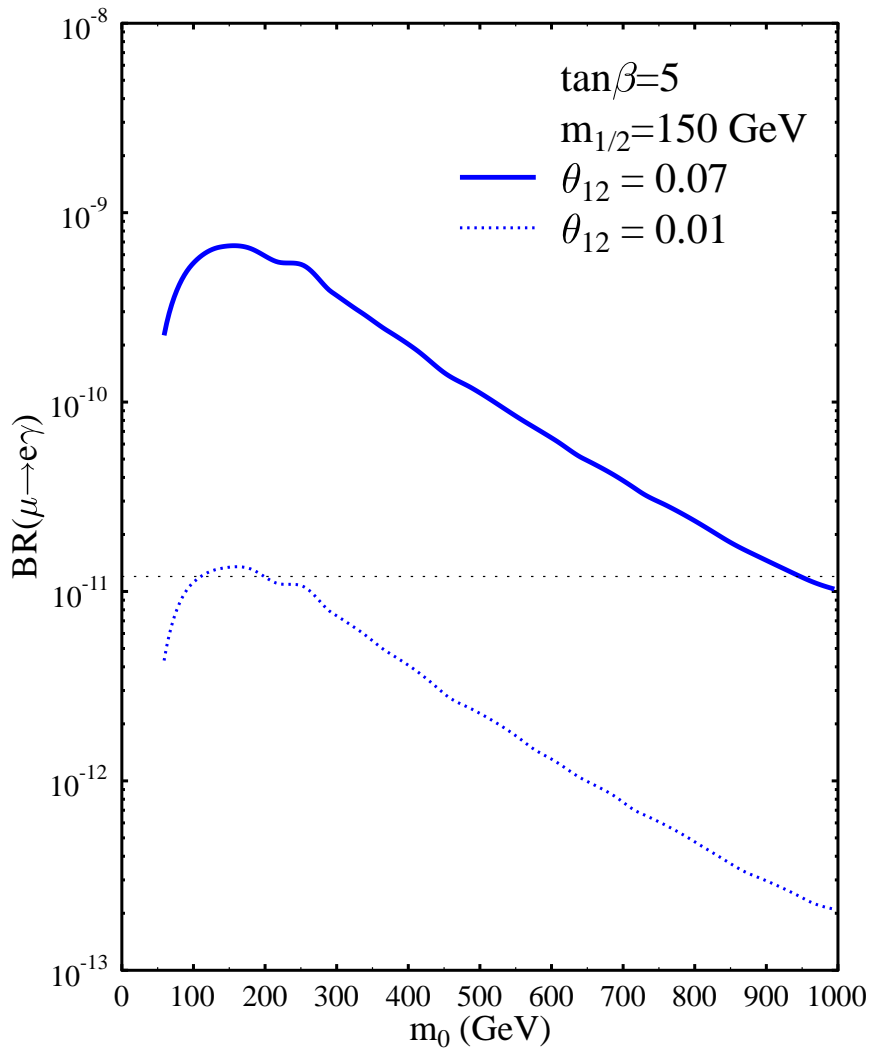}
\caption{\label{fig3} Branching ratio of $\mu\to e\gamma$ as a function of 
$m_0$ for $\tan\beta=5$, $m_{1/2}=150 GeV$. $A_0=m_0$ and
$\mu > 0$ are assumed.
$\theta_{12}$ is the mixing angle between the 1-2 generations defined in
Eq. (\ref{angles}). The horizontal dotted line
is the present experimental limit for $\mu\to e\gamma$, $1.2\times 10^{-11}$\cite{mueg}. }
\end{figure}

In FIG. \ref{fig3} $Br(\mu\to e\gamma)$ is plotted as a function of $m_0$ for
$\tan\beta=5$, with $m_{1/2}=150 GeV$. $\theta_{12}$ represents
 the 1-2 mixing angle defined in Eq. (\ref{angles}) (We omit the bar over
$\theta$ thereafter).  We notice that if $\theta_{12}$ takes the
typical value 0.07 in the ``lopsided'' models the predicted branching ratio 
is far larger than the present experimental limit in most parameter space.
For $\tan\beta=5$ we can suppress the branching
ratio below the experimental limit when we take $\theta_{12}$ as small as
0.01. 

\begin{figure}
\includegraphics[scale=0.8]{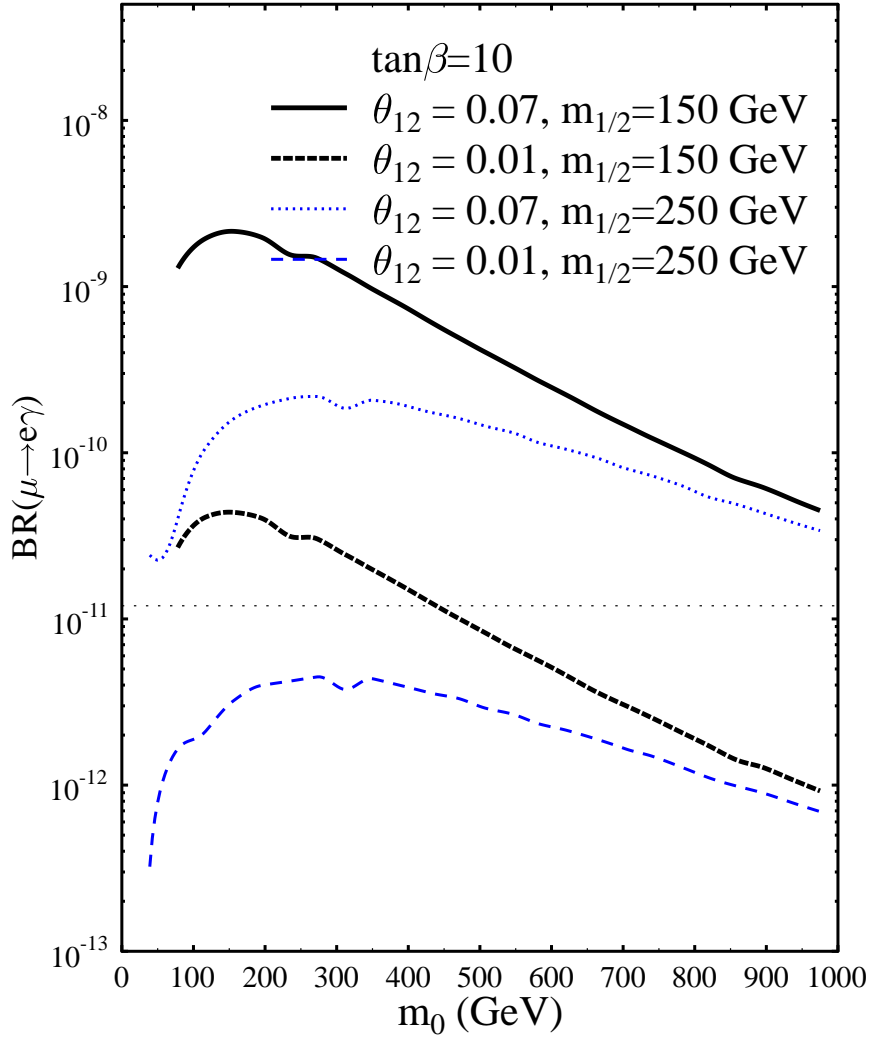}
\caption{\label{fig4} Branching ratio of $\mu\to e\gamma$ as a function of
$m_0$ for $\tan\beta=10$ and $m_{1/2}=150 GeV, 250 GeV$. 
Other comments are the same as that given in
FIG. \ref{fig3}. 
} \end{figure}

FIG. \ref{fig4} plots $Br(\mu\to e\gamma)$ as a function of $m_0$ for
$\tan\beta=10$ and $m_{1/2}=150,\ 250 GeV$ respectively. 
For the typical value $\theta_{12}=0.07$ the predicted branching ratio is always larger than
the experimental limit. 
In case of $m_{1/2}=150 GeV$, even we
take $\theta_{12}=0.01$ the predicted branching ratio is still too large for 
$m_0 \lesssim 470 GeV$, where $\delta a_\mu$ falls within the $1\sigma$ range.
When $m_{1/2}$ is $250 GeV$, the branching ratio can be below the experimental
limit when $\theta_{12}=0.01$.

\begin{figure}
\includegraphics[scale=0.6]{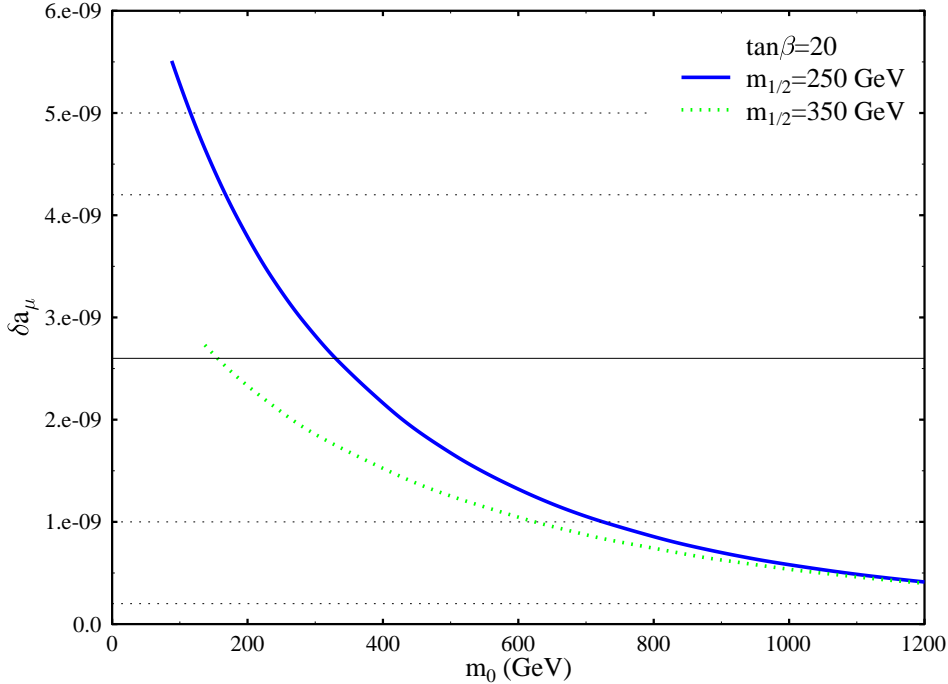}
\caption{\label{fig6} $\delta a_\mu$ as a function of $m_0$ for $\tan\beta=20$, 
$m_{1/2}=250 GeV, 350 GeV$ respectively. Other comments are the same as that 
given in FIG. \ref{fig2}. }
\end{figure}

We display $\delta a_\mu$ as a function of $m_0$ for $\tan\beta=20$
in FIG. \ref{fig6}, with $m_{1/2}=250, 350 GeV$ respectively.
The value of $m_0$ in the $\pm 1\sigma$ region can now be as large
as about $700 GeV$.
%We notice that the two curves do not separate much.
%According to Eq. (\ref{rela}) we expect the corresponding $Br(\mu\to e\gamma)$
%do not separate much too for the two cases $m_{1/2}=250 GeV$ and 
%$m_{1/2}=350 GeV$. We find
%this is true by our explicit calculations. So we only present the figures
%for the case of $m_{1/2}=250 GeV$ below. 

\begin{figure}
\includegraphics[scale=0.8]{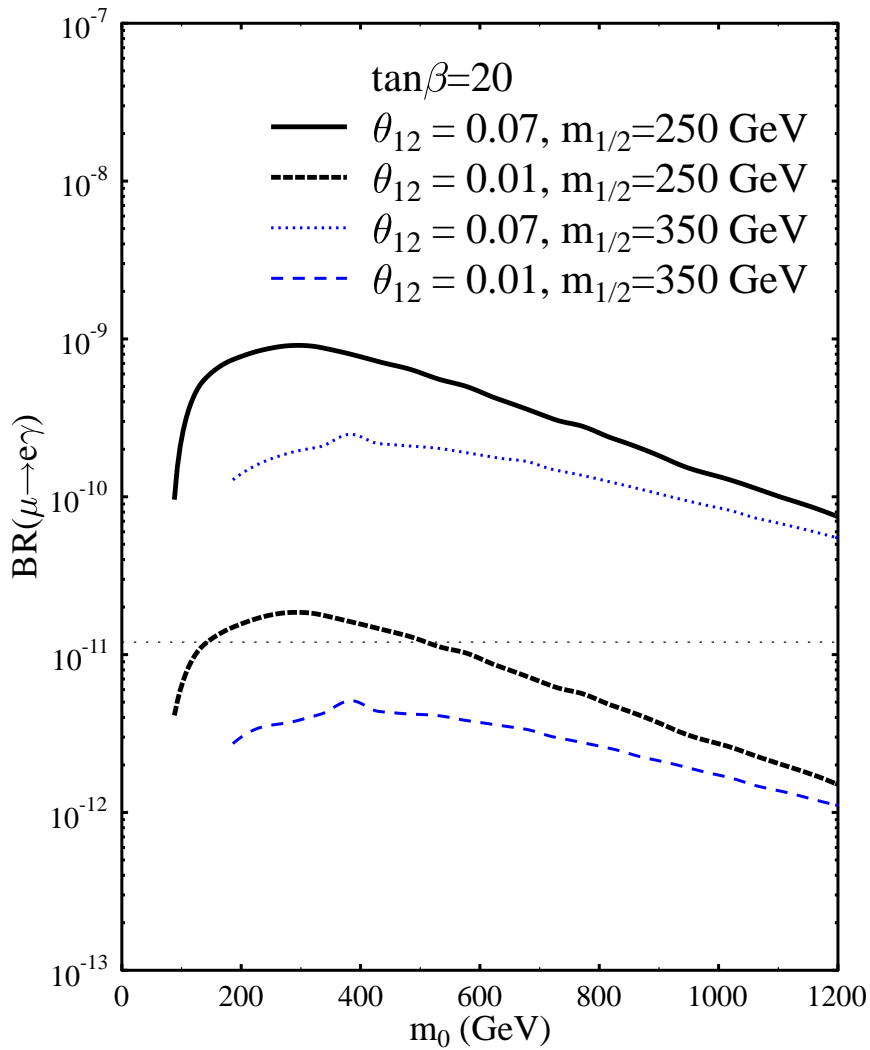}
\caption{\label{fig7} Branching ratio of $\mu\to e\gamma$ as a function of 
$m_0$ for $\tan\beta=20$, $m_{1/2}=250 GeV, 350 GeV$.
Other comments are the same as that given in FIG. \ref{fig3}. }
\end{figure}

FIG. \ref{fig7} is similar to FIG. \ref{fig4} except that 
$\tan\beta=20$ and $m_{1/2}=250 GeV$ and $350 GeV$ now.
In case of $\tan\beta=20$, we must
take $m_{1/2}$ as large as $350 GeV$ so that we can get a suppressed 
$Br(\mu\to e\gamma)$ when $\theta_{12}=0.01$ in the $\pm 1\sigma$ region of
$\delta a_\mu$. 

In summary, the branching ratio of $\mu\to e\gamma$ is sensitive to
the SUSY parameters $m_0$, $m_{1/2}$ and $\tan\beta$. For the typical value
$\theta_{12}=0.07$, in most parameter space
%allowed by the muon anomalous magnetic moment, 
the ``lopsided'' models
predict a $Br(\mu\to e\gamma)$ greater than the present experimental limit.
Only when $\theta_{12}$ is as small as $0.01$ can the predicted branching ratio
be below the present limit. Following this consideration, we build a new 
kind of models which can produce a near maximal $2-3$ mixing, large $1-2$ mixing 
in the charged lepton sector and extremely small $({V_D})_{13}$.

\section{An interesting form of charged lepton mass matrix}

We have shown in the last section that the ``lopsided'' models usually
predict a larger branching ratio of $\mu\to e\gamma$ than the present
experimental limit. The large $2-3$ generation
mixing of the charged leptons
enhances both the elements $({V_D})_{23}$ and $({V_D})_{13}$. Furthermore,
the scale of $M_R$ is lower than that in the case when the large neutrino 
mixing comes from $M_\nu$. 
Several authors noticed this fact and pointed out that this might
imply a universal condition for the soft-SUSY  breaking terms at the
energy scale not much higher than the weak scale\cite{casas,sato}, 
such as in the gauge mediated SUSY breaking (GMSB) models. 

If we do not give up the attractive supergravity models and at the same time
insist that the maximal atmospheric neutrino mixing is mainly coming from the
charged lepton mixing, the sole way to suppress the process $\mu\to e\gamma$
is to suppress $({V_D})_{13}$. We have examined many such models with 
symmetric elements between the first row and the first column in the charged 
lepton mass matrix. We found that $\bar{s}_{12}$ is always given by 
$\sqrt{m_e/m_\mu}\sim 0.07$ with a coefficient of order 
$1$. So the model with such structure will generally give a too large
branching ratio of $\mu\to e\gamma$, % than the present experimental limit,
as shown in the last section. 
However, we find $({V_D})_{13}$ can be greatly suppressed in a new form of the
charged lepton mass matrix with the elements of $1-3$ generations 
asymmetric. The amazing thing of this kind of models is that a large
$1-3$ element in $M_L$ can lead to a large $1-2$ generation mixing, 
which leads to the LMA solution to the solar neutrino problem, and a very small
$({V_D})_{13}$ at the same time. 
%So, contrary to the conclusion of \cite{casas}, the LMA solusion does not imply a large $Br(\mu\to e\gamma)$. 
As shown in Ref. \cite{barr},
the LMA solution is usually difficult to be constructed and most neutrino mass
models predict a SMA or VO solution for the solar neutrinos. The reason
for this is that most models try to produce both the large $1-2$ mixing, with 
$\tan^2\theta_{sol}$ in the range from about 0.2 to 0.8, 
and the small ratio of mass-squared split 
$\Delta m^2_{sol}/\Delta 
m^2_{atm}\approx 1.4\times 10^{-2}$ in the neutrino mass matrix $M_\nu$.
This is hard to be achieved and sometimes fine tuning is needed. 
However, if large $\theta_{sol}$ is produced
in the charged lepton sector it will be very simple to obtain the required
neutrino  spectrum in $M_\nu$. 

In the basis where the Dirac neutrino mass matrix is diagonal 
we give the charged lepton mass matrix as
\be
\label{new}
M_L\ \sim\ \left( \begin{array}{ccc}
0 & \delta & \sigma' \\ -\delta & 0 & 1-\epsilon \\ 0 & \epsilon & 1 \end{array} \right) \ ,%\ \ M_D\ \sim \ \left( \begin{array}{ccc} 0 & -\delta  & \epsilon' \\ \delta & 0 & -\epsilon/3 \\ \sigma' & 1 & 1 \end{array} \right) \ ,
\ene
with $\delta=0.00075$, $\sigma'=0.6$ and $\epsilon=0.12$.
$\epsilon$ is chosen to fit $m_\mu/m_\tau$ and $\delta$ to fit $m_e/m_\mu$.
$\sigma'$ can always suppress the value of $({V_D})_{13}$ and at the same time
predict a large $({V_D})_{12}$. We then get
$\tan^2\theta_{12}=0.45$, $\sin^22\theta_{23}=0.997$ and $({V_D})_{13}=0.0052$.
These values are approximately the corresponding values in MNS matrix,
since if we assume $Y_N$ has a similar hierarchical structure to that of up quark
it will transfer to $M_\nu$ and give very small mixing in $U_\nu$\cite{barr}.
% if we take $M_R$ approximately diagonal. 
If we take $\sigma'$ to fit the 
value of $\tan^2\theta_{12}$ we have two predictions as $({V_D})_{13}=0.0052$ and
$\theta_{23}\approx \pi/4$. We can also take $\tan^2\theta_{12}$ and
$\theta_{23}\approx \pi/4$ as two predictions and smallness of $({V_D})_{13}$
as the requirement. With a large $\sigma'$, a small $({V_D})_{13}$ and large 
$({V_D})_{12}$ are produced naturally without any fine tuning.

It should be noted that we only want to show an interesting form of the charged 
lepton mass matrix, like Eq. (\ref{new}), which can suppress the $1-3$ mixing and
enhance $1-2$ mixing in $V_D$. We do not intend to build a complete
model here and the quark sector is not discussed. Even though, it should be noticed that a large $\sigma'$ will not lead to a large mixing in the down quarks.
%We will dicuss the complete model which
%predicts all the quark and lepton masses and mixing in detail eleswhere. 
%However, it seems in such a complet model the requirement of 
%small $Br(\mu\to e\gamma)$ is hard to achieve again.

\begin{figure}
\includegraphics[scale=0.8]{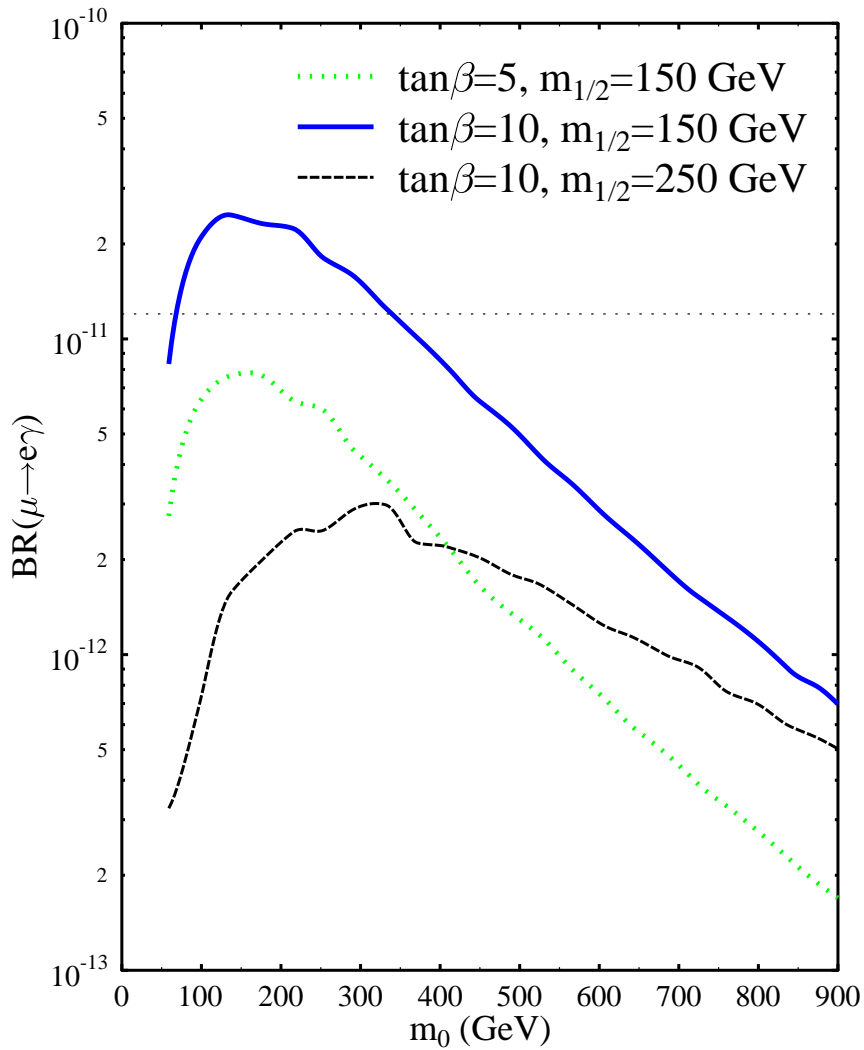}
\caption{\label{newfig3} Branching ratio of $\mu\to e\gamma$ predicted by our
model is plotted as a function 
of $m_0$ for $\tan\beta=5$, $m_{1/2}=150 GeV$ and $\tan\beta=10$, 
$m_{1/2}=150 GeV, 250 GeV$ respectively.
Other comments are the same as that given in FIG. \ref{fig3}. }
\end{figure}

\begin{figure}
\includegraphics[scale=0.8]{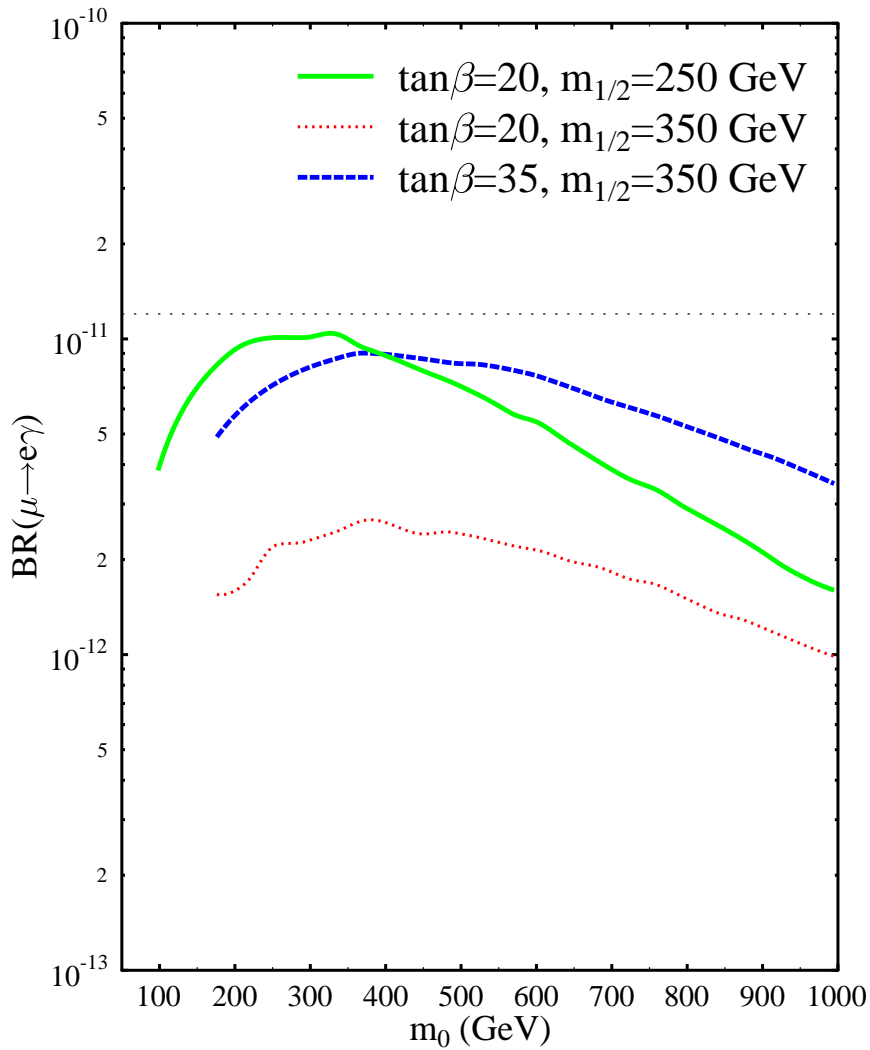}
\caption{\label{newfig4} Branching ratio of $\mu\to e\gamma$ predicted by our
model is plotted as a function of $m_0$ for $\tan\beta=20$, 
$m_{1/2}=250 GeV, 350 GeV$ and for $\tan\beta=35$, $m_{1/2}=350 GeV$ respectively.
Other comments are the same as that given in FIG. \ref{fig3}. }
\end{figure}

In FIG. \ref{newfig3},  we show the branching ratio of $\mu\to e\gamma$ predicted
by the above model with $\tan\beta=5, 10$. 
FIG. \ref{newfig4} shows the predicted branching ratio of the model in case of
$\tan\beta=20$ and $\tan\beta=35$. 
In most parameter space in these figures the decay $\mu\to e\gamma$ has 
a branching ratio
close to the present experimental limit and can be easily discovered by the
future experiments.
%By adjusting the parameter $\sigma'$ in Eq. (\ref{new}) larger 
%we get smaller branching ratio of $\mu\to e\gamma$. 
%However, in that case $\tan^2\theta_{sol}$ 
%will be larger than the present best fitted value. 

\section{conclusion and discussion}

In this work, we calculate the branching ratio of $\mu\to e\gamma$ in the
neutrino mass models with ``lopsided'' texture of the charged lepton mass matrix, 
within the supersymmetric framework. The constraints set on the supersymmetric
parameter space by the muon anomalous magnetic moment are considered. 
If the charged lepton mass matrix has
symmetric elements between the $1-2$ and $1-3$ generations,
this kind of models will generally predict a much larger branching ratio 
of $\mu\to e\gamma$ than the present experimental limit in most SUSY 
parameter space.  

To accommodate the experimental data, strong constraint should be 
satisfied by the charged lepton mass matrix, 
\textit{i.e.}, $({V_D})_{13}\lesssim 0.01$. 
We then show a new kind of models with asymmetric elements between the
$1-3$ generations, in addition to the asymmetry between the $2-3$ generations
in the ``lopsided'' models. The large $1-3$ element of the charged lepton mass
matrix can suppress $({V_D})_{13}$ and at the same time predict a large 
mixing between the $1-2$ generations naturally. 
This feature is most interesting in constructing the LMA solution to the solar 
neutrino problem, which is usually hard to be achieved in the 
literature\cite{barr}. 

With the improvement of the sensitivity to the LFV processes by three or more orders in the 
next generation experiments\cite{nexp}, it is quite possible that the LFV 
processes will be discovered in
the near future. $\tau\to\mu\gamma$ is a promising process to determine
whether there is a large mixing between the $2-3$ generations in the charged
lepton sector, such as in the ``lopsided'' models\cite{bi}. 
If both $\tau\to\mu\gamma$ and $\mu\to e\gamma$ are discovered in the near 
future, a kind of lepton structure as given by the present work should be 
very attractive.
If only $\tau\to \mu\gamma$ is found, this will imply a
large mixing between the $2-3$ generations in the charged lepton sector. 
However, very special structure should be designed to suppress 
the $\mu\to e\gamma$ process further. On the contrary, if $\tau\to \mu\gamma$ is 
not found while $\mu\to e\gamma$ is found, this should be easy to understand
and in most models without a lopsided structure this case is predicted. 
The last possibility is that no LFV is found at all.  In this case 
the most natural explanation is that the soft-SUSY breaking parameters
are universal at the energy scale not much higher than the weak scale, such as
in the GMSB mechanism. 

If LFV is found in the near future, the particles inducing LFV, such as the 
SUSY particles, may be produced directly on the high 
energy colliders. The measurements on high energy colliders will help us
to infer the flavor structure when combining with the LFV experimental results.
At the same time, interplay with the experiments of neutrino oscillation
is important. If LFV is found and at the same time $U_{e3}$ of the MNS matrix is
measured, much more constraint is obtained in building the neutrino mass
models. In any case, the measurement of lepton flavor violation is quite 
important to infer the
possible lepton flavor structure or other possible new physics, 
no matter the result is positive or null. 

\begin{acknowledgments}
This work is supported by the National Natural Science Foundation 
of China under the grant No. 10105004 and No. 19835060.
\end{acknowledgments}

%\bibliography{lfv}

\end{document}